\documentclass[12pt]{article}
\date{Feb 8, 2025}
\usepackage[a4paper,width=160mm,top=25mm,bottom=25mm]{geometry}
\usepackage{amsmath}
\usepackage{esint}
\usepackage{parskip}
\usepackage{physics}
\usepackage{stix}
\usepackage{bm}
\usepackage{hyperref}
\usepackage{mathtools}
\usepackage{comment}
\usepackage[sort,compress,numbers]{natbib}
\usepackage{authblk}
\newcommand*\hhbar{\hbar_0}
\title{The reformed hydrodynamic Schrödinger equation and
 the effective method for eliminating pseudo-quantum potential: a case study on barotropic potential flow[V2]}
\author[1]{Yi-Sian Ciou}
\numberwithin{equation}{section}
\begin{document}
\maketitle
\begin{abstract}
Given the discrepancies between the framework of the hydrodynamic Schrödinger equation (HSE) and classical fluid dynamics, we propose a reformed framework, termed the reformed hydrodynamic Schrödinger equation (RHSE) for clarity. 
The RHSE for barotropic potential flow is developed; notably,
    we demonstrate an alternative approach to eliminate the pseudo-quantum potential without resorting to the classical limit.
We call this approach the “correction method.” We then examine the correction method using Noether's theorem,
    confirming that it enforces the conservation of energy and momentum. Furthermore,
    we find that the classical limit fails to completely remove redundant terms when a two-component wave function is used to introduce vorticity. 
This fact suggests that the classical limit is not a universal fallback,
    highlighting the importance of further investigating the correction method.
\end{abstract}
\subsection*{Announcement}
This is the version of the article before peer review or editing, as submitted by an author to \emph{Fluid Dynamics Research}.
 IOP Publishing Ltd is not responsible for any errors or omissions in this version of the manuscript or any version derived from it.
 The Version of Record is available online at [INSERT DOI].
 The author welcomes constructive feedback and potential collaboration with experts to further refine and enhance this work.
\section{Introduction}
    Quantum computing is not limited to simulations of quantum systems\ \cite{Nielsen_Chuang_2010};
    its application to classical fluid systems has also been explored (e.g.,\ \cite{yepez1998quantum,mezzacapo2015quantum}).
    However, due to the inherent linearity of quantum mechanics\ \cite{wilson2023origin},
    quantum computing is currently constrained to solving linear problems\ \cite{romina_yalovetzky__2024}.
    This limitation has prompted efforts to translate classical problems into forms compatible with quantum mechanics.
Through the Madelung transformation\ \cite{Madelung:1927aa},
the Schrödinger equations (SE) is reformulated into a pair of equations resembling those in fluid dynamics,
which is outlined below.
The SE reads
   \begin{equation}\label{Schrödinger equation}
       {i}\hbar\pdv{\psi}{t}=\widehat{H}\psi,\;
       \widehat{H} \equiv -\frac{\hbar^2}{2m}\laplacian+V,
   \end{equation}
where $\widehat{H}$ is the Hamiltonian operator, $m$ denotes particle mass, and $V$ represents potential energy.
The \emph{Madelung wave function}\ \cite{lopez2004nonlinear} is commonly written as $\psi(\vb{x},t)=\sqrt{\rho}\exp(iS/\hbar)$,
where $\rho(\vb{x},t)$ denotes the probability density of finding a particle at position $\vb{x}$\ \cite[p.57]{greiner1996fieldQuanti},
and $S(\vb{x},t)$ represents Hamilton's principal function\ \cite[p.96]{sakurai2020modern}. 
Expressing $\psi$ in terms of its amplitude and phase, the SE splits into
  \begin{equation}\label{Hamilton-Jacobi eq}
      \pdv{S}{t}+\frac{\abs{\grad{S}}^2}{2m} +V +Q =0,\;
      Q\equiv -\frac{\hbar^2}{2m}\frac{\laplacian(\sqrt{\rho})}{\sqrt{\rho}},
  \end{equation}
  \begin{equation}\label{continuity eq}
      \pdv{\rho}{t}+\divergence(\rho\vb{v})=0,
  \end{equation}
where $Q$ is a quantum potential, known as the \emph{Bohm potential}\ \cite{chern2017fluid,lopez2004nonlinear}, and $\vb{v}=(\grad{S})/m$.
The equations above are the Hamilton-Jacobi equation\ \cite{sakurai2020modern,ballentine2014quantum} and the continuity equation, respectively.
What is shown above is known as the Madelung transformation.
    Subsequently, this has led to the development of various derivative Schrödinger-like equations (e.g.,\ \cite{chavanis2017,dhaouadi2019,meng2023quantum,chern2017fluid,chern2016smoke}),
    collectively referred to as the family of hydrodynamic Schrödinger equation (HSE).
However, there are two main discrepancies between the HSE framework and classical fluid dynamics:
(1) The primary obstacle preventing HSEs from describing classical fluids is the presence of quantum potentials (i.e., terms involving the reduced Planck constant $\hbar$) embedded in their resultant equations of motion.
The most common approach for eliminating quantum potentials is to apply the classical limit\ \cite[p.96]{sakurai2020modern}, i.e., $\hbar\rightarrow 0$.
(2) HSEs explicitly depend on particle mass.
In fluid dynamics, matter is generally treated as a continuous medium rather than as discrete molecules,
    which highlights an inherent incompatibility with the current HSE framework.
Note that expressing $S/m$, $h/m$, and $V/m$ using different symbols may hide the explicit presence of particle mass,
but they still depend on it. Meanwhile, although $[S/m]$ shares the same dimension as velocity potential,
they are intrinsically different quantities. This implies that setting $m=1$ fails to disengage the dependence on particle mass as well.
Regarding the physical meaning of $\rho$, some researchers have substituted mass density for its original role as probability density (e.g.,\ \cite{meng2023quantum,chern2016smoke,chern2017fluid}).
Such a transition indeed brings the HSEs closer to classical fluid dynamics,
but it has also caused an issue: while $[\psi]$ changes to ${\rm M^{1/2}\rm L^{-3/2}}$, $[\widehat{H}]$ is not modified accordingly,
which prevents a corresponding functional for a given HSE from being a valid Lagrangian density\footnote{If one directly replaces the Schrödinger wave function with the Madelung wave function ($\rho$ denotes mass density) in the Lagrangian density of the SE (see Eq.\ (\ref{Lagrangian of SE})),
then the dimension of that functional no longer aligns with energy density.
Yet, varying it with respect to $\psi^*$ still produces a corresponding HSE.}.
This transition also results in dimensional inconsistency when calculating the expectation value of an observable,
i.e., $[\int\psi^* O\psi\dd[3]{\vb{x}}]=[O]{\rm M}$, where $O$ represents an observable.
These key issues motivate the development of a reformed HSE framework that better aligns with classical fluid dynamics.

    The structure of this paper is as follows. In Sec.\ \ref{Sec: Reform},
    we begin by modifying the Madelung wave function,
    after which the reformed hydrodynamic Schrödinger equation (RHSE) is formulated accordingly.
    Focusing on eliminating the redundant term resembling the Bohm potential,
    we derive the RHSE for barotropic potential flow and demonstrate how the correction method works.
    In Sec.\ \ref{Sec: L field theory},
    we derive the corresponding Lagrangian density of the RHSE obtained previously,
    and apply Noether's theorem to examine whether the correction method ensures the conservation of energy and momentum.
    Sec.\ \ref{Sec: Discussion} discusses the scenario in which the classical limit may fail.
    Lastly, Sec.\ \ref{Sec: Conclusion} concludes the paper with a summary of our findings and a discussion of future perspectives.

\section{The RHSE and the correction method}\label{Sec: Reform}
\subsection{The postulates of the RHSE}\label{Sec: dimension issue and reform}
An appropriate form of wave function should be written as
\begin{equation}\label{an appropriate form of wave function}
    \psi(\vb{x},t)=\sqrt{\rho} e^{i\phi/\hhbar},
\end{equation}
where $\rho$ represents mass density of fluid,
 and $\phi(\vb{x},t)$ denotes a non-uniquely defined velocity potential such that velocity is exactly expressed as $\grad{\phi}$ (detailed in Appendix\ \ref{Appendix: linear momentum density});
 $\hhbar$ is a new constant introduced to satisfy that the phase is dimensionless, i.e., $[\hhbar]=[\phi]$.
Additionally, to distinguish $\hhbar$ from $\hbar$,
we refer to $\hhbar$ as \emph{hydrodynamic Planck constant}. Upon these changes,
we propose the one-component RHSE
 \begin{equation}\label{reformed Schrödinger eq}
    i\hhbar\pdv{\psi}{t}=\widehat{\mathcal{H}}\psi,
\end{equation}
where $\widehat{\mathcal{H}}$ is defined as \emph{specific Hamiltonian operator}\footnote{The term ``specific'' is borrowed from thermodynamics, where energy per unit mass is referred to as ``specific energy''\ \cite[p.37]{graebel}.},
with a dimension of energy per unit mass. Similar to the Hamiltonian operator,
the specific Hamiltonian operator can be decomposed into
\begin{equation}\label{specific Hamiltonian operator}
    \widehat{\mathcal{H}}\equiv-\frac{\hhbar^2}{2}\laplacian{}+ w + \Phi,
\end{equation}
where $\Phi(\vb{x},t)$ denotes potential of an external force, typically gravity, in the context of fluid dynamics;
 $w$ represents specific enthalpy of fluid (i.e., enthalpy per unit mass)\ \cite[p.4]{landau2013fluid}.
Recall the postulate that RHSE should disengage the dependence on particle mass,
this means that $\hhbar$ is a mass-independent quantity.
Albeit $\hhbar$ and $\hbar$ are related by $[\hhbar]=[\hbar]{\rm M}^{-1}$,
they cannot be directly linked by $\hhbar=\hbar/m$,
as particle mass has no utility in this framework.
The RHSE is deliberately formulated without dependence on particle mass,
making it more ``hydrodynamic'' in nature. Lastly, concerning the value of $\hhbar$, it is a minor consideration,
as it simply approaches zero if one adopts the ``classical limit,\@'' which in this context means $\hhbar\rightarrow 0$. Alternatively,
 one may adopt the \emph{correction} method, allowing $\hhbar$ to take an arbitrary nonzero value,
  since no term involving $\hhbar$ appears in a resultant equation of motion, as will be discussed below.

\subsection{RHSE for barotropic potential flow}\label{Sec: RHSE for barotropic potential flow}
We see from Eq.\ (\ref{Hamilton-Jacobi eq}) that the Bohm potential arises from the variable probability density.
Likewise, in this framework, a structurally similar potential emerges due to compressibility of fluid.
 Given that such a potential is redundant,
we focus on developing the RHSE for compressible potential flow and explore an alternative approach to eliminate it beyond the classical limit.
At this juncture, we shall first briefly discuss the properties of potential flow.
Since potential flow is a consequence of Kelvin's theorem (i.e., the law of conservation of circulation),
which requires the assumption that the flow be isentropic\ \cite[p.16]{landau2013fluid},
it follows that for a compressible potential flow, there must be an injective relationship between mass density and pressure\ \cite[p.13]{landau2013fluid}.
In other words, such a fluid should be barotropic. Thus, for barotropic potential flow,
 mass density and pressure are linked by the polytropic equation of state\ \cite{cherubini2013classical,Filippi2018VonMises,cherubini2011VonMises}
\begin{equation}\label{polytropic eq of state}
    p=K\rho^\gamma,
\end{equation}
where $p$ denotes the pressure, $K$ is a constant, and $\gamma$ is the adiabatic index\ \cite[p.114]{blundell2006thermal}.
Under these conditions, the specific enthalpy for barotropic potential flow takes the form
\begin{equation}\label{specific enthalpy}
    w=\int\frac{\dd{p}}{\rho}=\frac{\gamma}{\gamma-1}\frac{p}{\rho}.
\end{equation}
Since $\rho=\abs{\psi}^2$, Eq.\ (\ref{specific enthalpy}) can be rewritten as 
\begin{equation}\label{specific enthalpy resembles nonlinear term in GPE}
    w=\frac{K\gamma}{\gamma-1}\abs{\psi}^{2(\gamma-1)},
\end{equation}
which resembles the nonlinear potential in the Gross-Pitaevskii equation (GPE)\ \cite[Eq. (1)]{perez1997GPE}.

With the relations established above, we can immediately write down a basic form of the RHSE for barotropic potential flow
\begin{equation}\label{RHSE for barotropic without external potential}
    i\hhbar\pdv{\psi}{t}=\left(
        -\frac{\hhbar^2}{2}\laplacian + \frac{\gamma}{\gamma-1}\frac{p}{\rho}
    \right)\psi,
\end{equation}
where the potential $\Phi$ is ignored for simplicity,
and this simplification is maintained throughout the following discussion.
By expressing $\psi$ in terms of its amplitude and phase, Eq.\ (\ref{RHSE for barotropic without external potential}) splits into
\begin{equation}\label{rescaled:imaginary part}
    i\hhbar\pdv{(\sqrt{\rho})}{t}=-\frac{i\hhbar}{2}\left(
        2\grad{\phi}\cdot\grad(\sqrt{\rho})+\sqrt{\rho}\laplacian{\phi}
    \right),
\end{equation}
and
\begin{equation}\label{rescaled:real part}
    -\sqrt{\rho}\pdv{\phi}{t}=-\frac{\hhbar^2}{2}\left(
        \laplacian(\sqrt{\rho})-\frac{1}{\hhbar^2}\sqrt{\rho}\abs{\grad{\phi}}^2
    \right)+\frac{\gamma}{\gamma-1}\frac{p}{\rho}\sqrt{\rho}.
\end{equation}
Multiplying Eq.\ (\ref{rescaled:imaginary part}) with $2\sqrt{\rho}/i\hhbar$ and replacing $\grad{\phi}$ with $\vb{v}$,
we obtain the continuity equation.
Dividing Eq.\ (\ref{rescaled:real part}) by $-\sqrt{\rho}$ leads to
\begin{equation}\label{Bernoulli eq involving Bohm potential}
    \pdv{\phi}{t}+\frac{1}{2}\abs{\grad{\phi}}^2+\frac{\gamma}{\gamma-1}\frac{p}{\rho}
    -\frac{\hhbar^2}{2}\frac{\laplacian(\sqrt{\rho})}{\sqrt{\rho}}=0,
\end{equation}
which closely resembles the Bernoulli equation but involves a \emph{pseudo}-quantum potential.
Note that we add a prefix ``pseudo-'' to distinguish the potential involving $\hhbar$ from real quantum potentials involving $\hbar$,
since this framework is no longer quantum mechanical in nature.
The pseudo-quantum potential originates from the gap between this framework and quantum mechanics,
conceptually similar to a “vestigial structure” inherited from quantum mechanics.
To eliminate such a redundant potential, other than the classical limit,
we may also consider eliminating it by incorporating the negative pseudo-quantum potential into the specific Hamiltonian operator, namely,
\begin{equation}\label{potential for barotropic flow}
    \widehat{\mathcal{H}}:=-\frac{\hhbar^2}{2}\laplacian + \frac{\gamma}{\gamma-1}\frac{p}{\rho}
    +\frac{\hhbar^2}{2}\frac{\laplacian(\sqrt{\rho})}{\sqrt{\rho}}.
\end{equation}
Accordingly, Eq.\ (\ref{RHSE for barotropic without external potential}) becomes
\begin{equation}\label{RHSE for barotropic}
    i\hhbar\pdv{\psi}{t}=\left(
        -\frac{\hhbar^2}{2}\laplacian+\frac{\gamma}{\gamma-1}\frac{p}{\rho}
        +\frac{\hhbar^2}{2}\frac{\laplacian(\sqrt{\rho})}{\sqrt{\rho}}
    \right)\psi,
\end{equation}
which leads to
\begin{equation}\label{Bernoulli eq for barotropic potential flow}
    \pdv{\phi}{t}+\frac{1}{2}\abs{\grad{\phi}}^2+\frac{\gamma}{\gamma-1}\frac{p}{\rho}=0.
\end{equation}
This result matches the Bernoulli equation for barotropic potential flow\ \cite{cherubini2013classical,Filippi2018VonMises,cherubini2011VonMises}.
Subsequently, taking gradient of Eq.\ (\ref{Bernoulli eq for barotropic potential flow}) and replacing $\grad{\phi}$ by $\vb{v}$,
we obtain the Euler equation
\begin{equation}
    \pdv{\vb{v}}{t}+\vb{v}\cdot\grad{\vb{v}}=-\frac{1}{\rho}\grad{p},
\end{equation}
where
\begin{equation}
    \grad(\frac{\gamma}{\gamma-1}\frac{p}{\rho})=\frac{K}{\rho}\gamma\rho^{\gamma-1}\grad{\rho}
    =\frac{1}{\rho}\grad{p}.
\end{equation}
Incidentally, for incompressible potential flow, the corresponding RHSE is simply
\begin{equation}
    i\hhbar\pdv{\psi}{t}=\left(
        -\frac{\hhbar^2}{2}\laplacian+\frac{p}{\rho_0}
    \right)\psi,
\end{equation}
where $\rho_0$ denotes the constant mass density.

Thus far, the correction method—namely, introducing the negative pseudo-quantum potential into the RHSE—appears to provide
 a straightforward path to the Bernoulli equation for classical fluids\@.
However, from a field-theoretic perspective, this approach is somewhat puzzling: 
the negative pseudo-quantum potential, referred to here as the correction term,
 does not exist in the physical world but is artificially added to the RHSE\@.
This conceptual peculiarity motivates the examination into this method more thoroughly;
specifically, we investigate whether this approach breaks the conservation of energy and momentum,
as will be demonstrated in the next section.

\section{Lagrangian field theory}\label{Sec: L field theory}
\subsection{The effective Lagrangian density}
Resembing the structure of the SE, the RHSE may also have a corresponding Lagrangian density analogous to that of the SE\@.
However, the presence of the correction term can complicate the derivation,
because its counterpart is unknown and not easily deduced.
To address this challenge, we utilize the energy-momentum tensor from Noether's theorem to identify the counterpart of the correction term.
Lastly, the Euler-Lagrange equation (ELE) is applied to verify whether the assumed Lagrangian density reproduces the RHSE (see Eq.\ \ref{RHSE for barotropic}).
A brief review of the derivation of the ELE is provided in Appendix\ \ref{Appendix: the least action},
 whereas the derivation of energy-momentum tensor is omitted here due to the extensive algebraic manipulations required;
 interested readers are referred to the relevant literature\ \cite{cherubini2009lagrangian,greiner2000relativistic,ohanian2013gravitation}.
In this section, we do not elaborate further on these principles but instead apply them directly to streamline the demonstration below.

The Lagrangian density for the SE reads\ \cite{cherubini2009lagrangian,greiner2000relativistic}
\begin{equation}\label{Lagrangian of SE}
    \mathcal{L}=\frac{-\hbar^2}{2m}\grad{\psi}\cdot\grad{\psi^*}
    +\frac{i\hbar}{2}\left(
        \psi^*\pdv{\psi}{t}-\psi\pdv{\psi^*}{t}
    \right)-\psi^*{V}{\psi}.
\end{equation}
Inserting Eq.\ (\ref{Lagrangian of SE}) into the ELE for fields (see Eq.\ (\ref{ELE for fields})) leads to the SE and its complex conjugate after some algebra.
Next, in analogy to Eq.\ (\ref{Lagrangian of SE}), we assume an effective Lagrangian density
\begin{equation}\label{assumed Lagrangian}
    \mathcal{L}=
        \frac{-\hhbar^2}{2}\grad{\psi}\cdot\grad{\psi^*}
        +\frac{i\hhbar}{2}\left(
    \psi^*\pdv{\psi}{t}-\psi\pdv{\psi^*}{t}
    \right)-\psi^* \left(\frac{K}{\gamma-1}\abs{\psi}^{2(\gamma-1)}-G
    \right)\psi,
\end{equation}
where 
\begin{equation}\label{specific internal energy}
    \frac{K}{\gamma-1}\abs{\psi}^{2(\gamma-1)}=\frac{1}{\gamma-1}\frac{p}{\rho},
\end{equation}
is the specific internal energy (i.e., internal energy per unit mass)\ \cite{cherubini2013classical,von2012mathematical};
 ${G}$ is an unknown function introduced to eliminate the pseudo-quantum potential. 
To find ${G}$, we resort to the energy-momentum tensor
\begin{equation}\label{energy-momentum tensor}
    {T^\mu}_\nu= \pdv{\mathcal{L}}{(\pdv*{\psi_A}{x^\mu})}\pdv{\psi_A}{x^\nu}-\mathcal{L}{\delta^\mu}_\nu,\;\mu=0,\ldots,3,
\end{equation}
where we use the contravariant four-vector\footnote{Here we adopt the notation in\ \cite{cherubini2009lagrangian} where speed of light is set to 1 so that $x^0\equiv t$.} $x^\mu=(x^0,x^i)\equiv(t,x,y,z)$ with the indices $i$ running from 1 to 3\ \cite[p.3]{greiner2000relativistic},
${\delta^\mu}_\nu$ being the Kronecker symbol.
Note that $\psi_A$ represents a multiplet of scalar independent fields\ \cite[see Appendix B]{cherubini2009lagrangian}; in this context $\psi_A=(\psi,\psi^*)$.
The component ${T^0}_0$ gives the Hamiltonian density (i.e., total energy density)
\begin{equation}\label{T00}
   \begin{split}
    {T^0}_0&=\pdv{\mathcal{L}}{(\partial_t\psi)}\pdv{\psi}{t}
    +\pdv{\mathcal{L}}{(\partial_t\psi^*)}\pdv{\psi^*}{t}-\mathcal{L}
    =\frac{\hhbar^2}{2}\abs{\grad{\psi}}^2+\frac{K}{\gamma-1}\abs{\psi}^{2\gamma}-{G}\abs{\psi}^2\\
    &=\left(
        \frac{1}{2}\rho\abs{\vb{v}}^2+\frac{p}{\gamma-1}
    \right)+\frac{\hhbar^2}{8}\frac{\abs{\grad{\rho}}^2}{\rho}-{G}\abs{\psi}^2,
   \end{split}
\end{equation}
where the expression within the parentheses is the total energy density,
and the last two terms should cancel out. Hence, we assume
\begin{equation}\label{G function}
    {G}\abs{\psi}^2=\frac{\hhbar^2}{8}\frac{\abs{\grad{\rho}}^2}{\rho}.
\end{equation}
To verify this assumption, we substitute Eq.\ (\ref{G function}) into Eq.\ (\ref{assumed Lagrangian}),
and then insert Eq.\ (\ref{assumed Lagrangian}) into the ELE, which, after some algebra (detailed in Appendix\ \ref{Appendix: Lagrangian density}), yields Eq.\ (\ref{RHSE for barotropic}).
Finally, we derived the effective Lagrangian density for barotropic potential flow:
\begin{equation}\label{Lagrangian density for barotropic inviscid flow}
   \begin{split}
    \mathcal{L}=
    \frac{-\hhbar^2}{2}\grad{\psi}\cdot\grad{\psi^*}
    &+\frac{i\hhbar}{2}\left(
    \psi^*\pdv{\psi}{t}-\psi\pdv{\psi^*}{t}
    \right)\\ &-\psi^* \left(\frac{K}{\gamma-1}\abs{\psi}^{2(\gamma-1)}
   -\frac{\hhbar^2}{8}\frac{\abs{\grad(\abs{\psi}^2)}^2}{\abs{\psi}^4}
    \right)\psi.
   \end{split}
\end{equation}
Notably, by expressing $\psi$ in terms of its amplitude and phase,
 Eq.\ (\ref{Lagrangian density for barotropic inviscid flow}) becomes
\begin{equation}\label{L can be written as pressure}
    \mathcal{L}=-\rho\left(
    \frac{1}{2}\abs{\vb{v}}^2 + \pdv{\phi}{t}
\right)-\frac{p}{\gamma-1}=p.
\end{equation}
In fact, this surprising result has been discovered in\ \cite{cherubini2013classical,schakel1996effective}.
However, such an expression is not in a useful form for deriving the equation of motion;
Eq.\ (\ref{L can be written as pressure}) can be rewritten as\ \cite[Eq. (12)]{cherubini2013classical}
\begin{equation}\label{Lagrangian density without gravitation}
    \mathcal{L}=p=K{\left(
        \frac{1-\gamma}{K\gamma}
    \right)}^{\gamma/\gamma-1}{\left(
        \pdv{\phi}{t}+\frac{1}{2}\abs{\grad{\phi}}^2
    \right)}^{\gamma/\gamma-1},
\end{equation}
following an elaborate transformation (detailed in Appendix\ \ref{Appendix: L becomes p}).
Substituting Eq.\ (\ref{Lagrangian density without gravitation}) into the ELE results in a more complex equation of motion,
known as the Von Mises equation\ \cite{cherubini2013classical,cherubini2011VonMises,Filippi2018VonMises}.

\subsection{The examination on the conservation of energy and momentum}
Given that the divergence of the energy-momentum tensor is zero, i.e.,\ \cite[Eq. (15)]{cherubini2013classical}
\begin{equation}\label{divergence of energy-momentum tensor}
    \pdv{x^\mu}({T^\mu}_\nu)=0,
\end{equation}
it provides an analytical tool for verifying whether the correction ensures the conservation of classical energy and linear momentum in the flow.
We begin with examining the conservation of classical energy by $\partial_\mu{T^\mu}_0=0$.
The Hamiltonian density reads (see Eq.\ (\ref{T00}))
\begin{equation}
    {T^0}_0 = \frac{1}{2}\rho\abs{\vb{v}}^2 + \frac{p}{\gamma-1}.
\end{equation}
And the components of energy flux vector ${T^i}_0$ are
\begin{equation}\label{energy flux}
    {T^i}_0=\pdv{\mathcal{L}}{(\pdv*{\psi}{x^i})}\pdv{\psi}{t} + \pdv{\mathcal{L}}{(\pdv*{\psi^*}{x^i})}\pdv{\psi^*}{t},
\end{equation}
where (see Eq.\ (\ref{ELE to barotropic-space coordinates}))
\begin{subequations}
    \begin{equation}
        \pdv{\mathcal{L}}{(\pdv*{\psi}{x^i})}
        =-\frac{\hhbar^2}{2}\pdv{\psi^*}{x^i}+\frac{\hhbar^2}{4}\frac{\psi^*}{\rho}\pdv{\rho}{x^i},
    \end{equation}
    \begin{equation}
        \pdv{\mathcal{L}}{(\pdv*{\psi^*}{x^i})}
        =-\frac{\hhbar^2}{2}\pdv{\psi}{x^i}+\frac{\hhbar^2}{4}\frac{\psi}{\rho}\pdv{\rho}{x^i}.
    \end{equation}
\end{subequations}
With these substitutes, Eq.\ (\ref{energy flux}) becomes
\begin{equation}
    \begin{split}
        {T^i}_0 &=-\frac{\hhbar^2}{2}\left(
        \pdv{\psi^*}{x^i}\pdv{\psi}{t}+\pdv{\psi}{x^i}\pdv{\psi^*}{t}
        \right)+\frac{\hhbar^2}{4}\frac{1}{\rho}\pdv{\rho}{x^i}\left(
        \psi^*\pdv{\psi}{t} + \psi\pdv{\psi^*}{t}
        \right) \\
    &=-\frac{\hhbar^2}{2}\left(
        \frac{1}{2\rho}\pdv{\rho}{x^i}\pdv{\rho}{t} + \frac{2\rho}{\hhbar^2}\pdv{\phi}{x^i}\pdv{\phi}{t}
    \right)+\frac{\hhbar^2}{4}\frac{1}{\rho}\pdv{\rho}{x^i}\pdv{\rho}{t} \\
    &=-\rho v_i\pdv{\phi}{t}.
    \end{split}
\end{equation}
With Eq.\ (\ref{Bernoulli eq for barotropic potential flow}), we find
\begin{equation}
    {T^i}_0=\rho{v_i}\left(
    \frac{1}{2}\abs{\vb{v}}^2+\frac{\gamma}{\gamma-1}\frac{p}{\rho}
\right).
\end{equation}
Hence, we obtain\ \cite[cf. Eq. (6.1), p.10]{landau2013fluid}
\begin{equation}
    0= \pdv{x^\mu}({T^\mu}_0)
    = \pdv{t}(\frac{1}{2}\rho\abs{\vb{v}}^2 + \frac{p}{\gamma-1})
    +\divergence[\rho\vb{v}\left(
        \frac{1}{2}\abs{\vb{v}}^2 + \frac{\gamma}{\gamma-1}\frac{p}{\rho}
    \right)].
\end{equation}
Likewise, we verify the conservation of linear momentum by $\partial_\mu {T^\mu}_i=0$.
The components of linear momentum density (see Eq.\ (\ref{T(0i)=})) read
\begin{equation}
    {T^0}_i=-\rho\pdv{\phi}{x^i},
\end{equation}
and the components of momentum flux ${T^j}_i$ are
\begin{equation}
    \begin{split}
        {T^j}_i &=\pdv{\mathcal{L}}{(\pdv*{\psi}{x^j})}\pdv{\psi}{x^i}
        + \pdv{\mathcal{L}}{(\pdv*{\psi^*}{x^j})}\pdv{\psi^*}{x^i} -\mathcal{L}{\delta^j}_i \\
        &=-\frac{\hhbar^2}{2}\left(
            \pdv{\psi^*}{x^j}\pdv{\psi}{x^i} + \pdv{\psi}{x^j}\pdv{\psi^*}{x^i}
        \right) +\frac{\hhbar^2}{4}\frac{1}{\rho}\pdv{\rho}{x^j}\left(
            \psi^*\pdv{\psi}{x^i} + \psi\pdv{\psi^*}{x^i}
        \right) -p{\delta^j}_i\\ 
        &=-\rho\pdv{\phi}{x^j}\pdv{\phi}{x^i} -p{\delta^j}_i,
    \end{split}
\end{equation}
where $\mathcal{L}$ is replaced with $p$ by Eq.\ (\ref{L can be written as pressure}).
Consequently, we obtain\ \cite[p.11]{landau2013fluid}
\begin{equation}
    0=\pdv{x^\mu}({T^\mu}_i)=\pdv{t}(-\rho v_i) + \divergence(-\rho v_i \vb{v}) -\pdv{p}{x^i}.
\end{equation}
In sum, the correction does ensure the conservation of energy and momentum.
Note that if a barotropic potential flow occurs in a gravitational field,
gravity does not appear in the momentum equation derived from Noether's theorem.
This is because gravitational potential depends explicitly on spatial coordinates,
 which breaks the invariance of $\mathcal{L}$ under spatial translations\ \cite{cherubini2009lagrangian}.
 As a result, the law of conservation of linear momentum does not hold in this scenario.
 In contrast, the conservation of energy remains valid.

\section{Discussion}\label{Sec: Discussion}
To assess the significance of the correction method,
we compare the two approaches used to eliminate the pseudo-quantum potential.
One can observe that applying the classical limit also leads to the same results—namely,
 the Bernoulli equation for barotropic potential flow (see Eq.\ (\ref{Bernoulli eq for barotropic potential flow})) and the conservation of energy and momentum—without requiring complex calculations (not shown).
  At this juncture, it is worth considering whether the classical limit can serve as a universal fallback when vorticity enters the picture.
 In such a scenario, a two-component wave function is needed to introduce vorticity\ \cite{chern2016smoke,chern2017fluid,meng2023quantum}.
Specifically, take
\begin{equation}\label{two-component spinor}
    {\Psi}=\begin{pmatrix}
        \psi_1 \\ \psi_2
    \end{pmatrix}=\begin{pmatrix}
        \sqrt{\rho}\cos\theta e^{i\phi/\hhbar} \\ \sqrt{\rho}\sin\theta e^{i(\beta+\phi)/\hhbar}
    \end{pmatrix},
\end{equation}
where $\theta(\vb{x},t)$ and $\beta(\vb{x},t)$ are real-valued functions,
 and $\rho(\vb{x},t)=\abs{\psi_1}^2+\abs{\psi_2}^2$.
For simplicity, we consider an incompressible rotational flow whose effective Lagrangian density generally constructed as
\begin{equation}\label{basic form of L for a spinor}
    \begin{split}
        \mathcal{L}= &-\frac{\hhbar^2}{2}\left(
        \grad{\psi_1}\cdot\grad{\psi^*_1}+\grad{\psi_2}\cdot\grad{\psi^*_2}
    \right) + \frac{i\hhbar}{2}\left(
        \psi^*_1\pdv{\psi_1}{t}-\psi_1\pdv{\psi^*_1}{t}
    \right)\\ &+\frac{i\hhbar}{2}\left(
        \psi^*_2\pdv{\psi_2}{t}-\psi_2\pdv{\psi^*_2}{t}
    \right)
    -\mathcal{U},
    \end{split}
\end{equation}
where $\mathcal{U}$ is the internal energy density independent of $\pdv*{\psi_A}{t}$; here $\psi_A=(\psi_1,\psi^*_1,\psi_2,\psi^*_2)$.
In this way, the components of linear momentum ${T^0}_i$ are equal to
\begin{equation}
    {T^0}_i=-\rho_0\left(
        \pdv{\phi}{x^i} + \sin^2\theta \pdv{\beta}{x^i}
    \right),
\end{equation}
where the velocity exactly matches the Clebsch representation\ \cite{lamb1924hydrodynamics,Scholle2016PhysicsLettersA,scholle2015clebsch}, i.e., $\vb{v}\doteq\grad(\,)+(\,)\grad(\,)$.
Accordingly, the kinetic energy density is 
\begin{equation}
    \frac{1}{2}\rho_0\abs{\vb{v}}^2=\frac{1}{2}\rho_0\left(
        \abs{\grad{\phi}}^2 + \sin^4\theta\abs{\grad{\beta}}^2 +2\sin^2\theta\grad{\phi}\cdot\grad{\beta}
    \right).
\end{equation}
Yet, the ``pseudo-quantum kinetic energy density'' in Eq.\ (\ref{basic form of L for a spinor}) is 
\begin{equation}
    \frac{\hhbar^2}{2}\left(
        \abs{\grad{\psi_1}}^2 + \abs{\grad{\psi_2}}^2
    \right)=\rho_0\left[
       \frac{\hhbar^2}{2}\abs{\grad{\theta}}^2
    +\frac{1}{2}\abs{\vb{v}}^2 +\frac{\sin^2 (2\theta)}{8}\abs{\grad{\beta}}^2
    \right],
\end{equation}
where the first term on the right-hand side will vanish upon applying the classical limit but the third term still persists.
Clearly, the classical limit is not universal,
highlighting the importance of further investigating the correction method.

\section{Conclusion}\label{Sec: Conclusion}
In this paper, we propose a reformed framework of HSE termed as RHSE,
which demonstrates improved alignment with classical fluid dynamics. Moreover,
we discuss the approaches to eliminate the pseudo-quantum potential
and proved that the correction method ensures the conservation of energy and momentum.
Concerning the terminology, pseudo-quantum ``potential'' is so named because of its similarity of the Bohm potential.
Recall that the counterpart of the pseudo-quantum potential, $G$ (see Eq.\ (\ref{G function})), arises from the pseudo-quantum kinetic energy density, that is,
\begin{equation}\label{expanding quantum kinetic energy density}
    \frac{\hhbar^2}{2}\abs{\grad{\psi}}^2=\frac{1}{2}\rho\abs{\vb{v}}^2+\frac{\hhbar^2}{8}\frac{\abs{\grad{\rho}}^2}{\rho}.
\end{equation}
Designating such a term as potential is conceptually puzzling.
Although the physical significance of quantum potentials has been explored in the literature (e.g.,\ \cite{bohm1952,Bohm1952ll,schonberg1954,lopez2004nonlinear}),
we prefer to set it aside rather than propose a new interpretation,
as one of our primary goals is to eliminate any term involving $\hhbar$.
The correction method is effective; however, even under such a simplified fluid condition (barotropic and irrotational),
the RHSE remains highly nonlinear (see Eq.\ (\ref{RHSE for barotropic in terms of psi})).
Evidently, while the classical limit is preferable for barotropic potential flow,
it breaks down when a two-component wave function is used to introduce vorticity,
highlighting the importance of investigating whether the correction method remains effective in this scenario.
With regard to vorticity, the HSEs for rotational flows have been developed (e.g.,\ \cite{chern2017fluid,chern2016smoke,meng2023quantum}).
 Furthermore,
 the viscous term in the Navier-Stokes equation can be expressed in terms of Clebsch variables under the constraint of constant mass density,
as shown in\ \cite{scholle2015clebsch,Scholle2016PhysicsLettersA,Scholle2020water}.
Building on these works, it may be possible to derive the RHSE for classical incompressible viscous flow,
indicating the potential of this framework's development.

\section*{Acknowledgment}
The author thanks Professor Alessandro Rizzo and Professor Christian Cherubini for engaging in discussions on the material presented in this paper
and deeply appreciates their generosity in sharing their time and expertise.

\appendix
\section{The principle of least action}\label{Appendix: the least action}
It is beneficial to concisely outline the derivation of the ELE for fields.
Consider now a simply connected region $\Omega$, enclosed by a finite boundary $\partial\Omega$.
Recall that gravity was previously ignored, so $\mathcal{L}$ does not explicitly depend on spatial coordinates (i.e., $\pdv*{\mathcal{L}}{x^i}=0$)
and depends solely on $\psi_A$ and their first derivatives. The action $I$ is
\begin{equation}
    I=\int_{t_1}^{t_2}\dd{t}\int_{\Omega}\mathcal{L}(\psi_A,\partial_\mu \psi_A)\dd[3]{\vb{x}}.
\end{equation}
Varying the action with respect to $\psi_A$ gives (detailed algebraic manipulations are not shown)
\begin{equation}
   \begin{split}
    0=\variation{I}=\int_{t_1}^{t_2}\dd{t}
    \int_{\Omega} \variation{\psi_A}\left[
        \pdv{\mathcal{L}}{\psi_A} - \divergence(\pdv{\mathcal{L}}{(\grad{\psi_A})}) - \pdv{t}(\pdv{\mathcal{L}}{(\partial_t{\psi_A})})
    \right] \dd[3]{\vb{x}}\\ 
    +\int_{\Omega}\dd[3]{\vb{x}}{\left[
        \pdv{\mathcal{L}}{(\partial_t \psi_A)}\variation{\psi_A}
    \right]}_{t_1}^{t_2}
    + \int_{t_1}^{t_2}\dd{t}\int_{\Omega} \pdv{x^i}( \pdv{\mathcal{L}}{(\partial_i \psi_A)}\variation{\psi_A}) \dd[3]{\vb{x}}.
   \end{split}
\end{equation}
Following the usual manipulation\ \cite[p.481]{ohanian2013gravitation}, we impose $\variation{\psi_A}(t_1)=\variation{\psi_A}(t_2)=0$ and assume that $\variation{\psi_A}=0$ at the respective upper and lower limits of $x^i$.
 In this way, the integrals in the second line vanish. At this point,
 we shall interpret explicitly the meaning of the assumption.
Recall that in this study $\psi_A=(\psi,\psi^*)=\sqrt{\rho}\exp(\pm i\phi/\hhbar)$, so
\begin{equation}
    \variation{\psi_A}=\pdv{\psi_A}{\rho}\variation{\rho}+\pdv{\psi_A}{\phi}\variation{\phi},
\end{equation}
which means that both $\variation{\rho}$ and $\variation{\phi}$ are assumed to vanish at the boundary.
Note that this assumption does not require $\rho$ and $\phi$ to be zero or uniform at the boundary.
In contrast, if one imposes these conditions, no flow occurs in the region.
Returning to the derivation, due to the arbitrarily varied $\variation{\psi_A}$, the expression within the brackets in the first line must vanish,
 yielding the ELE for fields:
\begin{equation}\label{ELE for fields}
    0=\pdv{\mathcal{L}}{\psi_A} - \divergence(\pdv{\mathcal{L}}{(\grad{\psi_A})}) - \pdv{t}(\pdv{\mathcal{L}}{(\partial_t{\psi_A})}).
\end{equation}
Note that even though $\psi$ and $\psi^*$ share the same quantities ($\rho$ and $\phi$), they must be varied independently\ \cite[p.18]{greiner2000relativistic}.

\section{The linear momentum density}\label{Appendix: linear momentum density}
For the sake of rigor, this appendix demonstrates the reform yields the correct linear momentum density.
In fact, to derive the linear momentum density, a known RHSE under specific fluid condition is required,
analogous to how calculating the probability current needs the known SE\ \cite[p.29-30]{griffiths2019introduction}.

Recalling the continuity equation (see Eq.\ (\ref{continuity eq})),
we expand $\pdv*{\rho}{t}$ to give
\begin{equation}
    \pdv{\rho}{t}=\psi\pdv{\psi^*}{t}+\psi^*\pdv{\psi}{t}.
\end{equation}
Replacing $\pdv*{\psi}{t}$ and $\pdv*{\psi^*}{t}$ by Eq.\ (\ref{reformed Schrödinger eq}) and its complex conjugate,
we have
\begin{equation}
    \begin{split}
        \pdv{\rho}{t}&=\psi\left[
            \frac{-1}{i\hhbar}\left(
                -\frac{\hhbar^2}{2}\laplacian+w+\Phi
            \right)\psi^*
        \right]+\psi^*\left[
            \frac{1}{i\hhbar}\left(
                -\frac{\hhbar^2}{2}\laplacian+w+\Phi
            \right)\psi
        \right]\\
    &=\frac{\hhbar}{2i}\left(
        \psi\laplacian{\psi^*}-\psi^*\laplacian{\psi}
    \right).
    \end{split}
\end{equation}
Consider
\begin{subequations}
    \begin{equation}
        \psi\laplacian{\psi^*}=
        \divergence(\psi\grad{\psi^*})-\grad{\psi^*}\cdot\grad{\psi},
    \end{equation}
    \begin{equation}
        \psi^*\laplacian{\psi}=
        \divergence(\psi^*\grad{\psi})-\grad{\psi}\cdot\grad{\psi^*}.
    \end{equation}
\end{subequations}
With these changes, the continuity equation becomes
\begin{equation}
    0=\frac{\hhbar}{2i} \divergence(\psi\grad{\psi^*}-\psi^*\grad{\psi})
    +\divergence{\vb{J}},
\end{equation}
where
\begin{subequations}
    \begin{equation}
        \grad{\psi}=e^{i\phi/{\hhbar}}\left(
            \grad(\sqrt{\rho})+\frac{i}{\hhbar}\sqrt{\rho}\grad{\phi}
        \right),
    \end{equation}
    \begin{equation}
        \grad{\psi^*}=e^{-i\phi/{\hhbar}}\left(
            \grad(\sqrt{\rho})-\frac{i}{\hhbar}\sqrt{\rho}\grad{\phi}
        \right).
    \end{equation}
\end{subequations}
Consequently,
\begin{equation}\label{redefined momentum density}
    \vb{J}=\frac{\hhbar}{2i} \left(
    \psi^*\grad{\psi}-\psi\grad{\psi^*}
\right)=\rho\grad{\phi}.
\end{equation}
Incidentally, using the momentum-energy tensor also leads to the same result, i.e.,
\begin{equation}\label{T(0i)=}
    {T^0}_i =\pdv{\mathcal{L}}{(\partial_t\psi)}\pdv{\psi}{x^i}
        +\pdv{\mathcal{L}}{(\partial_t\psi^*)}\pdv{\psi^*}{x^i}
    =-\rho \pdv{\phi}{x^i},
\end{equation}
which correspond to the \emph{negative} components of linear momentum density.

\section{From the Lagrangian density to the RHSE}\label{Appendix: Lagrangian density}
Recall the assumed Lagrangian density (cf. Eq.\ (\ref{assumed Lagrangian}))
\begin{equation}
    \mathcal{L}=
        \frac{-\hhbar^2}{2}\grad{\psi}\cdot\grad{\psi^*}+\frac{i\hhbar}{2}\left(
        \psi^*\pdv{\psi}{t}-\psi\pdv{\psi^*}{t}
    \right)-\frac{K}{\gamma-1}\abs{\psi}^{2\gamma}
        +\frac{\hhbar^2}{8}\frac{\abs{\grad{\rho}}^2}{\rho}.
\end{equation}
Note that we have replaced ${G}\abs{\psi}^2$ by $\hhbar^2\abs{\grad{\rho}}^2/8\rho$ (see Eq.\ (\ref{G function})).
Calculate
\begin{equation}
    \pdv{\mathcal{L}}{\psi^*}=\frac{i\hhbar}{2}\pdv{\psi}{t}-
    \left(
        \frac{K\gamma}{\gamma-1}\abs{\psi}^{2(\gamma-1)}
    \right)\psi+\pdv{\psi^*}(\frac{\hhbar^2}{8}\frac{\abs{\grad{\rho}}^2}{\rho}),
\end{equation}
where
\begin{equation}
    \begin{split}
        \pdv{\psi^*}(\frac{\hhbar^2}{8}\frac{\abs{\grad{\rho}}^2}{\rho})
        &=\frac{\hhbar^2}{8}\left(
            -\frac{\abs{\grad{\rho}}^2}{\rho^2}\psi+\frac{1}{\rho}\pdv{(\grad{\rho}\cdot\grad{\rho})}{\psi^*}
        \right)\\
        &=-\frac{\hhbar^2}{8}\frac{\abs{\grad{\rho}}^2}{\rho^2}\psi
        +\frac{\hhbar^2}{4}\frac{(\grad{\rho}\cdot\grad{\psi})}{\rho},
    \end{split}
\end{equation}
by
\begin{equation}
    \grad{\rho}=\grad(\psi\psi^*)=\psi\grad{\psi^*}+\psi^*\grad{\psi}.
\end{equation}
For $x^0$:
\begin{equation}\label{ELE to barotropic-time}
    \pdv{\mathcal{L}}{(\partial_t\psi^*)}=-\frac{i\hhbar}{2}\psi,\;
    \pdv{t}(\pdv{\mathcal{L}}{(\partial_t\psi^*)})=-\frac{i\hhbar}{2}\pdv{\psi}{t}.
\end{equation}
For $x^i$:
\begin{equation}\label{ELE to barotropic-space coordinates}
    \begin{split}
        \pdv{\mathcal{L}}{(\grad{\psi^*})} &=
    -\frac{\hhbar^2}{2}\grad{\psi} + \pdv{(\grad{\psi^*})}(\frac{\hhbar^2}{8}\frac{\abs{\grad{\rho}}^2}{\rho}) \\
    &=-\frac{\hhbar^2}{2}\grad{\psi} + \frac{\hhbar^2}{4}\frac{1}{\rho}\psi\grad{\rho}.
    \end{split}
\end{equation}
Accordingly,
\begin{equation}
    \divergence(\pdv{\mathcal{L}}{(\grad{\psi^*})})
    =-\frac{\hhbar^2}{2}\laplacian{\psi}+\divergence(\frac{\hhbar^2}{4}\frac{\psi}{\rho}\grad{\rho}),
\end{equation}
where
\begin{equation}
    \begin{split}
        \divergence(\frac{\hhbar^2}{4}\frac{\psi}{\rho}\grad{\rho})
        &=\frac{\hhbar^2}{4}\left[
            \grad(\frac{\psi}{\rho})\cdot\grad{\rho}+\frac{\psi}{\rho}\laplacian{\rho}
        \right]\\
        &=\frac{\hhbar^2}{4}\left(
            \frac{\grad{\rho}\cdot\grad{\psi}}{\rho}-\frac{\abs{\grad{\rho}}^2}{\rho^2}\psi
        \right)+\frac{\hhbar^2}{4}\frac{\laplacian{\rho}}{\rho}\psi.
    \end{split}
\end{equation}
Finally, we obtain
\begin{equation}\label{the final step to RHSE for barotropic}
    \begin{split}
        0&=\pdv{\mathcal{L}}{\psi^*}-\divergence(\pdv{\mathcal{L}}{(\grad{\psi^*})})
            -\pdv{t}(\pdv{\mathcal{L}}{(\partial_t\psi^*)})\\
    &=i\hhbar\pdv{\psi}{t}
        -\left(-\frac{\hhbar^2}{2}\laplacian
            +\frac{K\gamma}{\gamma-1}\abs{\psi}^{2(\gamma-1)}
        \right)\psi-\frac{\hhbar^2}{4}\left(
            \frac{\laplacian{\rho}}{\rho}-\frac{\abs{\grad{\rho}}^2}{2\rho^2}
        \right)\psi.
    \end{split}
\end{equation}
Next, since
\begin{equation}
    \frac{K\gamma}{\gamma-1}\abs{\psi}^{2(\gamma-1)}
    =\frac{\gamma}{\gamma-1}\frac{K\rho^\gamma}{\rho}=\frac{\gamma}{\gamma-1}\frac{p}{\rho},
\end{equation}
and
\begin{equation}\label{expansion of the pseudo-quantum potential}
   \frac{\hhbar^2}{4}\left(
        \frac{\laplacian{\rho}}{\rho}-\frac{\abs{\grad{\rho}}^2}{2\rho^2}
    \right)=\frac{\hhbar^2}{2}\frac{\laplacian(\sqrt{\rho})}{\sqrt{\rho}},
\end{equation}
substituting these into Eq.\ (\ref{the final step to RHSE for barotropic}) leads to
\[0=i\hhbar\pdv{\psi}{t}
-\left(-\frac{\hhbar^2}{2}\laplacian
    +\frac{\gamma}{\gamma-1}\frac{p}{\rho}+\frac{\hhbar^2}{2}\frac{\laplacian(\sqrt{\rho})}{\sqrt{\rho}}
\right)\psi,\]
which reproduces Eq.\ (\ref{RHSE for barotropic}). To stress its nonlinearity, we can rewrite it as
\begin{equation}\label{RHSE for barotropic in terms of psi}
    i\hhbar\pdv{\psi}{t}=-\frac{\hhbar^2}{2}\laplacian{\psi}+\left[
        \frac{K\gamma}{\gamma-1}\abs{\psi}^{2(\gamma-1)}+\frac{\hhbar^2}{4}\left(
            \frac{\laplacian(\abs{\psi}^2)}{\abs{\psi}^2}-\frac{\abs{\grad(\abs{\psi}^2)}^2}{2\abs{\psi}^4}
        \right)
    \right]\psi.
\end{equation}

\section{The reformulation of the Lagrangian density}\label{Appendix: L becomes p}
We first rewrite Eq.\ (\ref{Bernoulli eq for barotropic potential flow}) as
\begin{equation}\label{rewritten Bernoulli eq for barotropic}
    \frac{\gamma}{\gamma-1}\frac{p}{\rho}=\frac{K\gamma\rho^{\gamma-1}}{\gamma-1}
    =-\left(
        \pdv{\phi}{t}+\frac{1}{2}\abs{\grad{\phi}}^2
    \right).
\end{equation}
Consider the local sound speed $c$ given by\ \cite[Eq. (12)]{cherubini2011VonMises}
\begin{equation}\label{sound speed}
    c^2=\dv{p}{\rho}=K\gamma\rho^{\gamma-1}.
\end{equation}
Substituting $K\gamma\rho^{\gamma-1}$ into Eq.\ (\ref{rewritten Bernoulli eq for barotropic}) results in\ \cite[Eq. (13)]{cherubini2011VonMises}
\begin{equation}\label{c squared}
    c^2=-(\gamma-1)\left(
    \pdv{\phi}{t}+\frac{1}{2}\abs{\grad{\phi}}^2
\right).
\end{equation}
We then rewrite Eq.\ (\ref{sound speed}) as
\begin{equation}
    \rho={\left(
        \frac{c^2}{K\gamma}
    \right)}^{\gamma-1}
    ={\left(
        \frac{1-\gamma}{K\gamma}
    \right)}^{\gamma-1}{\left(
        \pdv{\phi}{t}+\frac{1}{2}\abs{\grad{\phi}}^2
    \right)}^{\gamma-1},
\end{equation}
by Eq.\ (\ref{c squared}). Consequently,
the Lagrangian density (cf. Eq.\ (\ref{L can be written as pressure})) becomes\ \cite[Eq. (12)]{cherubini2013classical}
\begin{equation}
    \mathcal{L}=p=K{\left(
    \frac{1-\gamma}{K\gamma}
\right)}^{\gamma/\gamma-1}{\left(
    \pdv{\phi}{t}+\frac{1}{2}\abs{\grad{\phi}}^2
\right)}^{\gamma/\gamma-1}.
\end{equation}

\bibliography{citeHSE2}
\end{document}